\documentclass[english,showpacs,showkeys]{revtex4-2}
\usepackage[utf8]{inputenc}
\usepackage{color}
\usepackage{babel}
\usepackage{units}
\usepackage{bm}
\usepackage{amsmath}
\usepackage{amssymb}
\usepackage{graphicx}
\usepackage[bookmarks=true,bookmarksnumbered=false,bookmarksopen=true,bookmarksopenlevel=1,
 breaklinks=false,pdfborder={0 0 0},pdfborderstyle={},backref=false,colorlinks=true]
 {hyperref}
\hypersetup{pdftitle={Near-threshold resonance in e+e− → Λc Λ̄c process},
 pdfauthor={S. G. Salnikov},
 citecolor=blue,urlcolor=blue,linkcolor=blue}

\makeatletter

\providecommand{\tabularnewline}{\\}

\usepackage{bm}

\makeatother

\begin{document}
\title{Near-threshold resonance in $e^{+}e^{-}\rightarrow\Lambda_{c}\bar{\Lambda}_{c}$
process}

\author{S. G. Salnikov}
\email{S.G.Salnikov@inp.nsk.su}
\affiliation{Budker Institute of Nuclear Physics, 630090, Novosibirsk, Russia}
\affiliation{Novosibirsk State University, 630090, Novosibirsk, Russia}

\author{A. I. Milstein}
\email{A.I.Milstein@inp.nsk.su}
\affiliation{Budker Institute of Nuclear Physics, 630090, Novosibirsk, Russia}
\affiliation{Novosibirsk State University, 630090, Novosibirsk, Russia}

\date{\today}
\begin{abstract}
We discuss the influence of various contributions of the $\Lambda_{c}\bar{\Lambda}_{c}$
potential on the energy dependence of the cross section $e^{+}e^{-}\rightarrow\Lambda_{c}\bar{\Lambda}_{c}$
near the threshold. New BESIII experimental data on the cross section
and electromagnetic form factors $G_{E}$ and $G_{M}$ are taken into
account. Our predictions are in good agreement with experimental data.
We predict a bound state of $\Lambda_{c}\bar{\Lambda}_{c}$ at energy
$\sim\unit[38]{MeV}$ below the threshold that may manifest itself
in anomalous behavior of light meson production cross sections in
a given energy region.
\end{abstract}
\maketitle

\section{Introduction}

Currently, much attention has been drawn to the study of processes
in which hadrons are produced in $e^{+}e^{-}$ annihilation near the
threshold and a strong energy dependence of the cross sections is
manifested. For instance, these processes are %
\mbox{%
$e^{+}e^{-}\to p\bar{p}$%
}~\citep{Aubert2006,Lees2013,Lees2013a,Akhmetshin2016,Akhmetshin2019,Ablikim2019},
\mbox{%
$e^{+}e^{-}\to n\bar{n}$%
}~\citep{Achasov2014,Ablikim2021f,Achasov2022,Achasov2023a}, %
\mbox{%
$e^{+}e^{-}\to B\bar{B}$%
}~\citep{Aubert2009,Mizuk2021}, %
\mbox{%
$J/\psi(\psi')\to p\bar{p}\pi^{0}(\eta)$%
}~\citep{Bai2001,Ablikim2009,Bai2003}, %
\mbox{%
$J/\psi(\psi')\to p\bar{p}\omega(\gamma)$%
}~\citep{Bai2003,Ablikim2008,Alexander2010,Ablikim2012,Ablikim2013b},
\mbox{%
$e^{+}e^{-}\to\Lambda\bar{\Lambda}$%
}~\citep{Aubert2007,Ablikim2018,Ablikim2023}, %
\mbox{%
$e^{+}e^{-}\to\Lambda_{c}\bar{\Lambda}_{c}$%
}~\citep{Pakhlova2008,Ablikim2018b,Ablikim2023a}, and %
\mbox{%
$e^{+}e^{-}\to\phi\Lambda\bar{\Lambda}$%
}~\citep{Ablikim2021c}. The energy dependence of the corresponding
cross sections can be successfully explained by the interaction of
produced hadrons (the final-state interaction)~\citep{dmitriev2007final,dmitriev2014isoscalar,Dmitriev2016,Milstein2018,Milstein2022c,Haidenbauer2014,Kang2015,Dmitriev2016a,Milstein2017,Milstein2022,Milstein2022a,Salnikov2023,Milstein2021,Salnikov2023a,Haidenbauer2016,Haidenbauer2021,Haidenbauer2023a}.

The processes of hadroproduction near the threshold in $e^{+}e^{-}$
annihilation can be described as follows. First, a quark-antiquark
pair is produced at small distances $\sim1/Q$, where $Q$ is the
invariant mass of produced particles. Then at distances $\sim1/\Lambda_{QCD}$
a process of hadronization takes place. A system of produced hadrons
can be described by some wave function $\psi(r)$. Since the relative
speed of hadrons near the threshold is small, they interact with each
other for quite a long time. As a result, the wave function $\psi(r)$
differs significantly from that in the case of non-interacting hadrons.
In this picture, the amplitude $T$ of hadroproduction can be represented
in the form $T=T_{0}\cdot\psi(0)$, where $T_{0}$ is the amplitude
of quark-antiquark pair production at small distances, and $\psi(0)$
is the wave function of hadronic system at distances $r_{0}\lesssim1/\Lambda_{QCD}$.
Close to the threshold, a characteristic size of the wave function
is much larger than~$r_{0}$. The amplitude $T_{0}$ depends weakly
on energy near the threshold, while the function $\psi(0)$ has a
strong energy dependence. Thus, final-state interaction of hadrons
is responsible for the strong energy dependence of the cross section.
Note that the specific form of wave functions depends on quantum numbers
of produced particles (spin, isospin, orbital angular momentum, etc.).
However, the behavior of cross sections near the threshold has some
common features.

First of all, we note that the shapes of cross sections near the threshold
are approximated not by the usual Breit-Wigner formulas, but the Flatt\'{e}
formulas and their generalizations~\citep{Flatte1976}. These formulas
are applicable if there is either a loosely bound state or a virtual
state in a system of produced hadrons. In the first case there is
a bound state with the energy $\varepsilon<0$, $\left|\varepsilon\right|\ll\left|U\right|$,
where $U$ is characteristic value of potential (energy $\varepsilon$
is counted from the threshold of hadronic pair production). We'll
call this case a sub-threshold resonance. In the second case there
is no loosely bound state, but a slight increase of the depth of the
potential results in its appearance. In this case we will talk about
an above-threshold resonance. In both cases, the modulus of the scattering
length $a$ in the system of produced hadrons is much larger than
the characteristic size~$R$ of the potential. For a sub-threshold
resonance, we have $a>0$ and $\varepsilon=-1/Ma^{2}$. In the case
of a virtual state we have $a<0$, and the energy of the virtual state
by definition is $\varepsilon=1/Ma^{2}\ll\left|U\right|$ (here $M$
is the mass of produced hadron). A detailed discussion of this picture
can be found in Refs.~\citep{Salnikov2023,Salnikov2023a}.

The Flatt\'{e} formula is expressed through a small number of parameters
(a scattering length, an effective range of interaction)~\citep{Kalashnikova2019a}.
Therefore, to describe the near-threshold behavior of cross sections
we can use any potentials that reproduce the required values of these
parameters. The most convenient way to describe near-threshold resonances
is using a potential in its simplest form (for example, in the form
of rectangular potential well), finding the corresponding wave functions
and fitting the parameters of the potential to achieve the best agreement
with experimental data. Such approach makes it easy to take into account
the Coulomb interaction of produced charged particles, difference
in particle masses in the case of several channels and other specific
effects. This approach turned out to be especially convenient for
description of several coupled channels~\citep{Salnikov2023a}.

In this work we use our approach to describe the cross section of
$\Lambda_{c}\bar{\Lambda}_{c}$ pair production in $e^{+}e^{-}$ annihilation
near the threshold. Experimental data for the cross section of this
process were presented in Refs.~\citep{Pakhlova2008,Ablikim2018b,Ablikim2023a}.
In the first two papers the cross section of the process was measured,
but the data on the electromagnetic form factors were very limited.
In a recent paper~\citep{Ablikim2023a} experimental data for the
cross section of the process were obtained with much higher accuracy
than in~\citep{Pakhlova2008,Ablikim2018b}. Note that the data on
the cross sections in Refs.~\citep{Pakhlova2008,Ablikim2023a} differ
noticeably from each other. Also in Ref.~\citep{Ablikim2023a} the
values of electric form factor $G_{E}$ and magnetic form factor $G_{M}$
of $\Lambda_{c}$ were measured. The latter circumstance allows one
to reduce significantly the uncertainty of various contributions to
the $\Lambda_{c}\bar{\Lambda}_{c}$ interaction potential.

\section{Theoretical approach}

The approach to the description of the cross-section of process $e^{+}e^{-}\rightarrow\Lambda_{c}\bar{\Lambda}_{c}$
is similar to the case of $p\bar{p}$ and $n\bar{n}$ pair production
in $e^{+}e^{-}$ annihilation near the threshold (see Refs.~\citep{Milstein2018,Milstein2022c}
and references therein). However, the case of $\Lambda_{c}\bar{\Lambda}_{c}$
pair production is simpler than $p\bar{p}$ and $n\bar{n}$ production.
Firstly, $\Lambda_{c}\bar{\Lambda}_{c}$ is produced only in the isospin
state with $I=0$, in contrast to the case of nucleon-antinucleon
pair which can has isospin $I=0$ or $I=1$. In addition, for nucleon-antinucleon
pair it is necessary to take into account the isotopic invariance
violation (the proton and neutron mass difference and absence of Coulomb
interaction for $n\bar{n}$ pair). Also, for nucleon-antinucleon pair
it is necessary to take into account the noticeable imaginary part
of the optical potential, which takes into account the high probability
of pair annihilation into mesons. We have checked that in the case
of $\Lambda_{c}\bar{\Lambda}_{c}$ the potential can be considered
as a real quantity.

A pair $\Lambda_{c}\bar{\Lambda}_{c}$ produced in $e^{+}e^{-}$ annihilation
has quantum numbers $J^{PC}=1^{--}$ and the total spin of the system
is $S=1$, while the orbital angular momentum $l$ can be zero or
two due to the tensor forces. Thus, the interaction potential of $\Lambda_{c}$
and $\bar{\Lambda}_{c}$ can be written as ($\hbar=c=1$)
\begin{equation}
\mathcal{V}(r)=-\frac{\alpha}{r}+V_{S}(r)\,\delta_{l0}+\left(\frac{6}{Mr^{2}}+V_{D}(r)\right)\delta_{l2}+V_{T}(r)\,S_{12}\,.
\end{equation}
Here $\alpha$~is the fine-structure constant, $V_{S}(r)$ and $V_{D}(r)$
are the contributions to the central potentials in $S$-wave and $D$-wave,
respectively, $V_{T}(r)S_{12}$ is the tensor potential, $S_{12}=6\left(\bm{S}\cdot\bm{n}\right)^{2}-4$
is the tensor operator, $\bm{S}$~is the spin operator of $\Lambda_{c}\bar{\Lambda}_{c}$
pair, and~$\bm{n}=\bm{r}/r$. Note that $V_{D}(r)$ differs from
$V_{S}(r)$ due to spin-orbit interaction. Separating the angular
and radial variables, we obtain the equations for the radial part
$u(r)$ of wave function corresponding to the $S$-wave, and the radial
part $w(r)$ corresponding to the $D$-wave:
\begin{equation}
\left[\frac{p_{r}^{2}}{M}+\mathcal{V}(r)-E\right]\Psi(r)=0\,,\label{eq:schrodinger}
\end{equation}
where $M$~is the mass of $\Lambda_{c}$ baryon, $E$~is the energy
of a pair, counted from the threshold, and $\left(-p_{r}^{2}\right)$~is
the radial part of the Laplace operator. The wave function $\Psi(r)$
of the Schrödinger equation~(\ref{eq:schrodinger}) has two components,
namely, $\Psi(r)=\left(u(r),\,w(r)\right)^{T}$. In this basis, the
potential $\mathcal{V}(r)$ can be written in a matrix form
\begin{equation}
\mathcal{V}(r)=\begin{pmatrix}-\frac{\alpha}{r}+V_{S} & -2\sqrt{2}V_{T}\\
-2\sqrt{2}V_{T}\quad & -\frac{\alpha}{r}+\frac{6}{Mr^{2}}+V_{D}-2V_{T}
\end{pmatrix}.
\end{equation}

The Schrödinger equation~(\ref{eq:schrodinger}) has two linearly
independent solutions $\Psi_{1}(r)=\left(u_{1}(r),\,w_{1}(r)\right)^{T}$
and $\Psi_{2}(r)=\left(u_{2}(r),\,w_{2}(r)\right)^{T}$, having different
asymptotic behavior at large distances, see~\citep{Milstein2022a}
for more details. Electromagnetic form factors of $\Lambda_{c}$ are
expressed through these solutions as follows
\begin{align}
 & G_{E}=\mathcal{G}\left(u_{1}(0)-\sqrt{2}\,u_{2}(0)\right),\nonumber \\
 & G_{M}=\mathcal{G}\left(u_{1}(0)+\frac{1}{\sqrt{2}}u_{2}(0)\right).
\end{align}
Here $\mathcal{G}$~is the amplitude of $\Lambda_{c}\bar{\Lambda}_{c}$
pair production at small distances. Near the threshold we can consider
$\mathcal{G}$ to be independent of energy. However, in order to describe
experimental data in a wider energy region, it is convenient to represent
$\mathcal{G}$ in the form $\mathcal{G}=\mathcal{G}_{0}\cdot F_{D}(Q)$,
where $\mathcal{G}_{0}$~is a constant, and the dipole form factor
$F_{D}(Q)$ reads
\begin{equation}
F_{D}(Q)=\frac{1}{\left(1-\frac{Q^{2}}{Q_{0}^{2}}\right)^{2}}\,,\qquad Q=2M+E\,,\qquad Q_{0}=\unit[1]{GeV}\,.
\end{equation}
It is seen that the ratio $G_{E}/G_{M}$ is independent of $\mathcal{G}$
and equals
\begin{equation}
\frac{G_{E}}{G_{M}}=\frac{u_{1}(0)-\sqrt{2}\,u_{2}(0)}{u_{1}(0)+\frac{1}{\sqrt{2}}u_{2}(0)}\,.
\end{equation}
Thus, the ratio $G_{E}/G_{M}$ differs from unity only due to the
contribution of $D$-wave arising due to tensor forces. At threshold
the contribution of $D$-wave is zero so that $G_{E}=G_{M}$. The
integrated cross section of $\Lambda_{c}\bar{\Lambda}_{c}$ pair production
has the form
\begin{equation}
\sigma=\frac{\pi k\alpha^{2}}{2M^{3}}\left|\mathcal{G}\right|^{2}\left(\left|u_{1}(0)\right|^{2}+\left|u_{2}(0)\right|^{2}\right),\label{eq:sigma}
\end{equation}
and its strong energy dependence is determined by the functions $\left|u_{1}(0)\right|$
and $\left|u_{2}(0)\right|$.

\tabcolsep=2ex

\begin{table}
\begin{centering}
\begin{tabular}{|l|c|c|c|}
\hline 
 & $V_{S}$ & $V_{D}$ & $V_{T}$\tabularnewline
\hline 
$\unit[U]{(MeV)}$ & $-1025$ & $-156$ & $-64$\tabularnewline
\hline 
$\unit[R]{(fm)}$ & $1.05$ & $1.99$ & $0.78$\tabularnewline
\hline 
\end{tabular}
\par\end{centering}
\caption{The parameters of the potential of $\Lambda_{c}\bar{\Lambda}_{c}$
interaction.}\label{tab:params}
\end{table}

\begin{figure}
\begin{centering}
\includegraphics[totalheight=5.4cm]{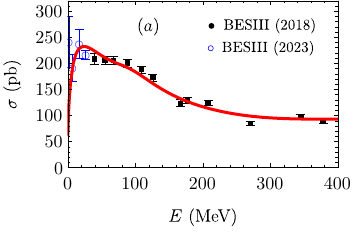}
\par\end{centering}
\begin{centering}
\includegraphics[totalheight=5.4cm]{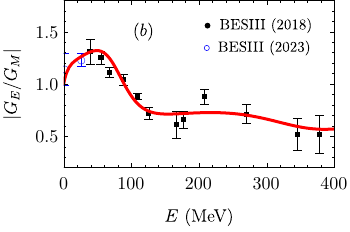}\hfill{}\includegraphics[totalheight=5.4cm]{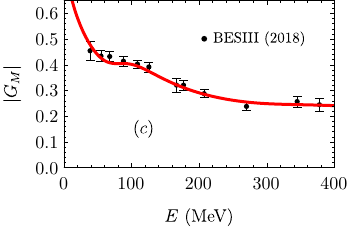}
\par\end{centering}
\caption{The cross section of the process $e^{+}e^{-}\to\Lambda_{c}\bar{\Lambda}_{c}$
in the energy region from the threshold to $\unit[400]{MeV}$, Fig.~(a).
The ratio of electric and magnetic form factors of $\Lambda_{c}$
baryon is shown in Fig.~(b) and the magnetic form factor is shown
in Fig.~(c). The curves correspond to the predictions of our model.
The experimental data are from Refs.~%
\mbox{%
\citep{Ablikim2018b,Ablikim2023a}%
}.}\label{fig:sigGeGm}
\end{figure}

It has been pointed out in the Introduction that, to describe near-threshold
behavior of the cross-section, one can use different shapes of potentials,
and the parameters of these potentials can be found by comparing theoretical
predictions with experimental data. In our work we choose these potentials
in the form of rectangular potential wells
\begin{equation}
V_{n}(r)=U_{n}\,\theta(R_{n}-r)\,,\qquad n=S,\,D,\,T\,,
\end{equation}
where $\theta(x)$~is the Heaviside function, $U_{n}$ and $R_{n}$~are
some fitting parameters. In addition, for convenience of numerical
calculations the tensor potential is regularized at small distances
by the factor
\begin{equation}
F(r)=\frac{(br)^{2}}{1+(br)^{2}}
\end{equation}
with $b=10\,\mbox{fm}^{-1}$. In fact, the results are almost independent
of the specific value of the parameter~$b$.

The parameters of potentials providing the best agreement with experimental
data~\citep{Ablikim2018b,Ablikim2023a} for the cross-section and
for the electromagnetic form factors $G_{E}$ and $G_{M}$ are given
in Table~\ref{tab:params}. Fig.~\ref{fig:sigGeGm}a shows a comparison
of the energy dependence of the cross section~(\ref{eq:sigma}) with
experimental data~\citep{Ablikim2018b,Ablikim2023a}. The energy
dependencies of $\left|G_{E}/G_{M}\right|$ and $\left|G_{M}\right|$
are shown in Figs.~\ref{fig:sigGeGm}b and~\ref{fig:sigGeGm}c,
respectively. It is seen that our prediction for $\sigma$, $\left|G_{E}/G_{M}\right|$
and $\left|G_{M}\right|$ are in good agreement with experimental
data. The corresponding ratio $\chi^{2}/N_{df}$ is $1.5$, where
$N_{df}$~is the number of degrees of freedom. For the parameters
specified in the Table~\ref{tab:params}, our model predicts a bound
state with energy $E_{0}=\unit[-38]{MeV}$. This bound state may manifest
itself in anomalous behavior of light meson production cross sections
in $e^{+}e^{-}$ annihilation near $E_{0}$ corresponding to $Q=\unit[4545]{MeV}$
(cf.~\citep{Milstein2022c} and references therein).

\section{Conclusion}

Using the latest BESIII data~\citep{Ablikim2018b,Ablikim2023a} on
the cross section of $\Lambda_{c}\bar{\Lambda}_{c}$ pair production
in $e^{+}e^{-}$ annihilation and corresponding electromagnetic form
factors $G_{E}$ and $G_{M}$, analysis of various contributions to
the $\Lambda_{c}$ and $\bar{\Lambda}_{c}$ interaction potential
was carried out. The consideration is based on an approach using effective
potential. Good agreement obtained of our predictions with experimental
data. Since the ratio $\left|G_{E}/G_{M}\right|$ differs significantly
from unity in a wide energy region, an account for tensor forces ($V_{T}$)
is crucially important, although $\left|V_{T}\right|\ll\left|V_{S}\right|$
(see Table~\ref{tab:params}). Note that an account for the potential
$V_{D}$, which differs from $V_{S}$ due to the spin-orbit interaction,
is also important. Influence of Coulomb interaction is noticeable
only in a narrow energy region near the threshold of $\Lambda_{c}\bar{\Lambda}_{c}$
production. We predict existence of a narrow sub-threshold resonance
in the system $\Lambda_{c}\bar{\Lambda}_{c}$ at energy $\sim\unit[38]{MeV}$
below the threshold. This bound state may manifest itself in anomalous
behavior of light meson production cross sections in a given energy
region.


\begin{thebibliography}{99}
	\bibitem{Aubert2006}
	B. Aubert, et al., \href{https://dx.doi.org/10.1103/PhysRevD.73.012005}{Phys. Rev. D \textbf{73}, 012005 (2006)}.
	\bibitem{Lees2013}
	J.P. Lees, et al., \href{https://dx.doi.org/10.1103/PhysRevD.87.092005}{Phys. Rev. D \textbf{87}, 092005 (2013)}.
	\bibitem{Lees2013a}
	J.P. Lees, et al., \href{https://dx.doi.org/10.1103/PhysRevD.88.072009}{Phys. Rev. D \textbf{88}, 072009 (2013)}.
	\bibitem{Akhmetshin2016}
	R.R. Akhmetshin, et al., \href{https://dx.doi.org/10.1016/j.physletb.2016.04.048}{Phys. Lett. B \textbf{759}, 634 (2016)}.
	\bibitem{Akhmetshin2019}
	R.R. Akhmetshin, et al., \href{https://dx.doi.org/10.1016/j.physletb.2019.05.032}{Phys. Lett. B \textbf{794}, 64 (2019)}.
	\bibitem{Ablikim2019}
	M. Ablikim, et al., \href{https://dx.doi.org/10.1103/PhysRevD.99.092002}{Phys. Rev. D \textbf{99}, 092002 (2019)}.
	\bibitem{Achasov2014}
	M.N. Achasov, et al., \href{https://dx.doi.org/10.1103/PhysRevD.90.112007}{Phys. Rev. D \textbf{90}, 112007 (2014)}.
	\bibitem{Ablikim2021f}
	M. Ablikim, et al., \href{https://dx.doi.org/10.1038/s41567-021-01345-6}{Nat. Phys. \textbf{17}, 1200 (2021)}.
	\bibitem{Achasov2022}
	M.N. Achasov, et al., \href{https://dx.doi.org/10.1140/epjc/s10052-022-10696-0}{Eur. Phys. J. C \textbf{82}, 761 (2022)}.
	\bibitem{Achasov2023a}
	M.N. Achasov, et al., \href{http://arxiv.org/abs/2309.05241}{arXiv:2309.05241 [hep-ex]}.
	\bibitem{Aubert2009}
	B. Aubert, et al., \href{https://dx.doi.org/10.1103/PhysRevLett.102.012001}{Phys. Rev. Lett. \textbf{102}, 012001 (2009)}.
	\bibitem{Mizuk2021}
	R. Mizuk, et al., \href{https://dx.doi.org/10.1007/JHEP06(2021)137}{J. High Energy Phys. \textbf{2021}, 137 (2021)}.
	\bibitem{Bai2001}
	J.Z. Bai, et al., \href{https://dx.doi.org/10.1016/S0370-2693(01)00605-0}{Phys. Lett. B \textbf{510}, 75 (2001)}.
	\bibitem{Ablikim2009}
	M. Ablikim, et al., \href{https://dx.doi.org/10.1103/PhysRevD.80.052004}{Phys. Rev. D \textbf{80}, 052004 (2009)}.
	\bibitem{Bai2003}
	J. Bai, et al., \href{https://dx.doi.org/10.1103/PhysRevLett.91.022001}{Phys. Rev. Lett. \textbf{91}, 022001 (2003)}.
	\bibitem{Ablikim2008}
	M. Ablikim, et al., \href{https://dx.doi.org/10.1140/epjc/s10052-007-0467-4}{Eur. Phys. J. C \textbf{53}, 15 (2008)}.
	\bibitem{Alexander2010}
	J. P. Alexander, et al., \href{https://dx.doi.org/10.1103/PhysRevD.82.092002}{Phys. Rev. D \textbf{82}, 092002 (2010)}.
	\bibitem{Ablikim2012}
	M. Ablikim, et al., \href{https://dx.doi.org/10.1103/PhysRevLett.108.112003}{Phys. Rev. Lett. \textbf{108}, 112003 (2012)}.
	\bibitem{Ablikim2013b}
	M. Ablikim, et al., \href{https://dx.doi.org/10.1103/PhysRevD.87.112004}{Phys. Rev. D \textbf{87}, 112004 (2013)}.
	\bibitem{Aubert2007}
	B. Aubert, et al., \href{https://dx.doi.org/10.1103/PhysRevD.76.092006}{Phys. Rev. D \textbf{76}, 092006 (2007)}.
	\bibitem{Ablikim2018}
	M. Ablikim, et al., \href{https://dx.doi.org/10.1103/PhysRevD.97.032013}{Phys. Rev. D \textbf{97}, 032013 (2018)}.
	\bibitem{Ablikim2023}
	M. Ablikim, et al., \href{https://dx.doi.org/10.1103/PhysRevD.107.072005}{Phys. Rev. D \textbf{107}, 072005 (2023)}.
	\bibitem{Pakhlova2008}
	G. Pakhlova, et al., \href{https://dx.doi.org/10.1103/PhysRevLett.101.172001}{Phys. Rev. Lett. \textbf{101}, 172001 (2008)}.
	\bibitem{Ablikim2018b}
	M. Ablikim, et al., \href{https://dx.doi.org/10.1103/PhysRevLett.120.132001}{Phys. Rev. Lett. \textbf{120}, 132001 (2018)}.
	\bibitem{Ablikim2023a}
	M. Ablikim, et al., \href{http://arxiv.org/abs/2307.07316}{arXiv:2307.07316 [hep-ex]}.
	\bibitem{Ablikim2021c}
	M. Ablikim, et al., \href{https://dx.doi.org/10.1103/PhysRevD.104.052006}{Phys. Rev. D \textbf{104}, 052006 (2021)}.
	\bibitem{dmitriev2007final}
	V.F. Dmitriev and A.I. Milstein, \href{https://dx.doi.org/10.1016/j.physletb.2007.06.085}{Phys. Lett. B \textbf{658}, 13 (2007)}.
	\bibitem{dmitriev2014isoscalar}
	V.F. Dmitriev, A.I. Milstein, and S.G. Salnikov, \href{https://dx.doi.org/10.1134/S1063778814080043}{Phys. At. Nucl. \textbf{77}, 1173 (2014)}.
	\bibitem{Dmitriev2016}
	V.F. Dmitriev, A.I. Milstein, and S.G. Salnikov, \href{https://dx.doi.org/10.1103/PhysRevD.93.034033}{Phys. Rev. D \textbf{93}, 034033 (2016)}.
	\bibitem{Milstein2018}
	A.I. Milstein and S.G. Salnikov, \href{https://dx.doi.org/10.1016/j.nuclphysa.2018.06.002}{Nucl. Phys. A \textbf{977}, 60 (2018)}.
	\bibitem{Milstein2022c}
	A.I. Milstein and S.G. Salnikov, \href{https://dx.doi.org/10.1103/PhysRevD.106.074012}{Phys. Rev. D \textbf{106}, 074012 (2022)}.
	\bibitem{Haidenbauer2014}
	J. Haidenbauer, X.-W. Kang, and U.-G. Meißner, \href{https://dx.doi.org/10.1016/j.nuclphysa.2014.06.007}{Nucl. Phys. A \textbf{929}, 102 (2014)}.
	\bibitem{Kang2015}
	X.-W. Kang, J. Haidenbauer, and U-G. Meißner, \href{https://dx.doi.org/10.1103/PhysRevD.91.074003}{Phys. Rev. D \textbf{91}, 074003 (2015)}.
	\bibitem{Dmitriev2016a}
	V.F. Dmitriev, A.I. Milstein, and S.G. Salnikov, \href{https://dx.doi.org/10.1016/j.physletb.2016.06.056}{Phys. Lett. B \textbf{760}, 139 (2016)}.
	\bibitem{Milstein2017}
	A.I. Milstein and S.G. Salnikov, \href{https://dx.doi.org/10.1016/j.nuclphysa.2017.06.002}{Nucl. Phys. A \textbf{966}, 54 (2017)}.
	\bibitem{Milstein2022}
	A.I. Milstein and S.G. Salnikov, \href{https://dx.doi.org/10.1103/PhysRevD.105.L031501}{Phys. Rev. D \textbf{105}, L031501 (2022)}.
	\bibitem{Milstein2022a}
	A.I. Milstein and S.G. Salnikov, \href{https://dx.doi.org/10.1103/PhysRevD.105.074002}{Phys. Rev. D \textbf{105}, 074002 (2022)}.
	\bibitem{Salnikov2023}
	A.I. Milstein and S.G. Salnikov, \href{https://dx.doi.org/10.1134/S0021364023601471}{JETP Lett. \textbf{117}, 901 (2023)}.
	\bibitem{Milstein2021}
	A.I. Milstein and S.G. Salnikov, \href{https://dx.doi.org/10.1103/PhysRevD.104.014007}{Phys. Rev. D \textbf{104}, 014007 (2021)}.
	\bibitem{Salnikov2023a}
	S.G. Salnikov, A.E. Bondar, and A.I. Milstein, \href{https://dx.doi.org/10.1016/j.nuclphysa.2023.122764}{Nucl. Phys. A, 122764 (2023)}.
	\bibitem{Haidenbauer2016}
	J. Haidenbauer and U.-G. Meißner, \href{https://dx.doi.org/10.1016/j.physletb.2016.08.067}{Phys. Lett. B \textbf{761}, 456 (2016)}.
	\bibitem{Haidenbauer2021}
	J. Haidenbauer, U.-G. Meißner, and L.-Y. Dai, \href{https://dx.doi.org/10.1103/PhysRevD.103.014028}{Phys. Rev. D \textbf{103}, 014028 (2021)}.
	\bibitem{Haidenbauer2023a}
	J. Haidenbauer and U-G. Meißner, \href{https://dx.doi.org/10.1140/epja/s10050-023-01017-4}{Eur. Phys. J. A \textbf{59}, 136 (2023)}.
	\bibitem{Flatte1976}
	S.M. Flatté, \href{https://dx.doi.org/10.1016/0370-2693(76)90654-7}{Phys. Lett. B \textbf{63}, 224 (1976)}.
	\bibitem{Kalashnikova2019a}
	Y. Kalashnikova and A. V. Nefediev, \href{https://dx.doi.org/10.3367/UFNe.2018.08.038411}{Physics-Uspekhi \textbf{62}, 568 (2019)}.
\end{thebibliography}
\end{document}